\documentclass[9pt,shortpaper,twoside,web]{ieeecolor2}
\usepackage{generic}
\usepackage{cite}
\usepackage{amsmath,amssymb,amsfonts}
\usepackage{algorithmic}
\usepackage{graphicx}
\usepackage{textcomp}
\usepackage{url}

\def\BibTeX{{\rm B\kern-.05em{\sc i\kern-.025em b}\kern-.08em
    T\kern-.1667em\lower.7ex\hbox{E}\kern-.125emX}}
\begin{document}
\title{\Huge A state-space framework for causal detection of hippocampal ripple-replay events}
\author{Sirui Zeng and Uri T. Eden
\thanks{This work was supported by research grants from the Simons Collaboration on the Global Brain (542971, NC-GB-CULM-00002730) and the NIH (RF1MH130623).}
\thanks{S. Zeng is with the Department of Mathematics and Statistics, Boston University, Boston, MA 02215, USA (e-mail: zengsr@bu.edu).}
\thanks{U. T. Eden is with the Department of Mathematics and Statistics, Boston University, Boston, MA 02215, USA (e-mail: tzvi@bu.edu).}
}

\maketitle

\begin{abstract}
Hippocampal ripple-replay events are typically identified using a two-step process that at each time point uses past and future data to determine whether an event is occurring. This prevents researchers from identifying these events in real time for closed-loop experiments. It also prevents the identification of periods of nonlocal representation that are not accompanied by large changes in the spectral content of the local field potentials (LFPs). In this work, we present a new state-space model framework that is able to detect concurrent changes in the rhythmic structure of LFPs with nonlocal activity in place cells to identify ripple-replay events in a causal manner. The model combines latent factors related to neural oscillations, represented space, and switches between coding properties to explain simultaneously the spiking activity from multiple units and the rhythmic content of LFPs recorded from multiple sources. The model is temporally causal, meaning that estimates of the switching state can be made at each instant using only past information from the spike and LFP signals, or can be combined with future data to refine those estimates. We applied this model framework to simulated and real hippocampal data to demonstrate its performance in identifying ripple-replay events. \end{abstract}


\section{Introduction}
\label{sec:introduction}
\IEEEPARstart{N}EUROSCIENCE experiments now routinely involve simultaneous recordings from large population of neurons \cite{b1}, \cite{b2}, \cite{b3}, \cite{b4}, and from multiple brain areas \cite{b5}, \cite{b6}. This has deepened our understanding of how different brain regions coordinate their activity to support complex cognitive processes. Technological advances have also allowed for the recording of neural activity at multiple spatial and temporal scales, which can provide a more comprehensive understanding of how neural systems interact to give rise to behavior and cognition. However, the increase in the size and dimensionality of neural datasets brings multiple new analysis challenges. One critical challenge is how to identify low-dimensional dynamics that represent internal cognitive processes from multisource electrophysiology data \cite{b7}, \cite{b8}. Additionally, neuroscience experiments increasingly include closed-loop designs, where neural signals are processed in real time and the outputs are used to guide features of the experiment \cite{b9}, \cite{b10}. 

These features are highlighted in the problem of identifying and decoding hippocampal ripple-replay events in rodents performing spatial navigation tasks \cite{b11}. During active exploration, hippocampal place cells fire selectively at specific locations in an environment and sequences of activity across the population reflect trajectories through that environment. Local field potentials (LFPs) exhibit a prominent theta rhythm (4-12 Hz) during periods of active exploration. When the animal is sleeping or resting, high frequency oscillations (150-250 Hz), called ripples, are observed in the LFPs \cite{b12}. Hippocampal ripples are often accompanied by replay, an organized reactivation of place cells that is interpreted as a trajectory through the environment despite the fact that the rodent is sleeping or at rest, and not moving through space \cite{b8}.

Despite the fact that ripple-replay events are characterized by distinct changes to both the LFPs and population spiking activity, current practice for detecting these events typically involves a two-stage process that first looks only at filtered local field potential (LFP) signals to identify candidate periods with prominent ripples, and subsequently examines the population spiking activity in these periods to identify nonlocal representations \cite{b13}, \cite{b14}. This two-stage process has a number of undesirable consequences. First, both stages of this procedure suffer from reduced statistical power relative to one that includes both the LFP and population spiking data, and there is the potential to exclude real events for lack of sufficient evidence at both stages. Second, this two-stage process leads to increased bias both in the identification of replay events (toward non-identification) and in the decoded content of each event, that is, the estimates of the nonlocal trajectories associated with each event. Third, this process is completely dependent on the assumption that ripples and replay always occur simultaneously, and would not be able to detect nonlocal spiking that does not co-occur with ripples. Fourth, this two-stage process is acausal in that it uses both past and future data to determine whether each time point is part of a ripple-replay event. Thus, this process is unsuited for any closed-loop experimental design that uses real-time estimates of replay state or replay content to determine whether and how to provide a stimulus. 

During replay events, sharp wave ripples (SWRs), composed of large amplitude events followed by fast oscillations between 150–200 Hz, are typically observed in the hippocampal LFPs. Detecting SWRs typically involves bandpass filtering the data in the ripple band, z-scoring the population power trace of an entire recording session, and finding times when the z-score exceeds a fixed threshold over a sustained time period. Typically, acausal filters are used for bandpass filtering and power estimation, precluding this approach from being used in online and closed-loop experiments. Additionally, spectral estimates are typically computed either from a single channel LFP or an averaged LFP, which limits the use of information from variability in LFP across channels in a brain region. 

Recently, a new latent-process modeling approach has been developed to decompose LFPs and other continuous electrophysiological signals into a small set of rhythmic components \cite{b15}, \cite{b16}. Importantly, this time-domain approach can be formulated so as to only use past information to estimate the state of each rhythmic component at the present time. This model framework has been recently extended to describe amplitude coupling analysis \cite{b17}, estimate instantaneous phase for multiple rhythms \cite{b18}, and model transitions in rhythmic functional connectivity patterns during induction into anesthesia \cite{b19}. However, these methods have so far been applied to only a handful of neural analysis problems for which all of the information relevant for estimation has been contained in the observed field data. Identifying ripple-replay events in real time requires expanding these methods to include information from the hippocampal population spiking activity.

Another critical challenge for real-time analysis of replay relates to the issue of spike sorting. Most decoding analyses of population spiking activity start by sorting the spikes into clusters based on their waveforms, fitting separate receptive field models to each cluster, and then decoding under the assumption that all the spikes have been correctly sorted \cite{b20}. Despite many advances in spike sorting in recent years \cite{b21}, \cite{b22}, \cite{b23}, \cite{b24}, spike sorting remains a major source of variability, induces bias in receptive field estimates \cite{b25}, and is not amenable to real-time analysis. Advances in neural recording technology now allow for recordings from hundreds to thousands of electrodes simultaneously, making systematic spike sorting less and less computationally tractable \cite{b26}, \cite{b27}, \cite{b28}.

Researchers have sought to eliminate the bias, variability, and computational challenge of spike sorting by developing clusterless, population-level neural coding models \cite{b13}, \cite{b14}, \cite{b29}, \cite{b30}, \cite{b31}, \cite{b32}, \cite{b33}. Instead of spike sorting and building separate receptive field models for each putative neuron, these methods forgo spike sorting and attempt to model the likelihood of observing a spike with a particular waveform or waveform feature from any of the neurons being recorded, as a function of the signals being encoded. Mathematically, this is a marked point process, where a random mark process describes the distribution of the spike waveform feature occurring for each spike \cite{b32}, \cite{b33}, \cite{b34}. Previously, we have shown that clusterless decoding of a rat's position from an ensemble of hippocampal place cells produces estimates that are more accurate and whose confidence regions are more correct than those from spike-sorted data \cite{b32}. Importantly, since they eliminate the need for spike-sorting, clusterless modeling methods can be performed in real time \cite{b55}.

Recent work has sought to integrate information from LFPs and population spiking to increase the statistical power of ripple-replay identification and content decoding \cite{b8}. This was accomplished using a hierarchical state-space model that included a set of discrete states describing the transitions between ripple-replay and non-ripple replay states and continuous states representing either true trajectories through space or the reactivation of trajectories that occur during replay. The observation models included a multivariate Gaussian model for the spectral power across multiple frequencies of the LFP and a set of point process models for the spiking of each neuron as a function of the rat's actual position or the latent spatial variable. This method could identify most of the potential replay events with relatively strong confidence and decode the content of those events. However, the spectral power estimates were non-causal, making this method unsuitable for real-time analysis and closed-loop experiments. 

To address these issues, we have developed a new hierarchical state-space model that includes a switching state to capture transitions to the ripple-replay state, two continuous latent processes that encode the spatial content of each replay and the state of a set of oscillators that describe the LFP rhythms, and two causal observation processes: one marked point process model for the population spiking data and one multivariate Gaussian process for the LFPs. We derive filter and smoother algorithms that simultaneously estimate the presence of a ripple-replay event, decode its spacial content, and decompose the LFPs into a small number of narrow-bandwidth rhythms. These algorithms provide posterior densities which allow us to compute estimates of each of these quantities as well as confidence regions, and allow us to perform specific hypothesis tests about the presence of events and their content. Thus this model is able to integrate information about ripple-replay events from both the spiking and LFP signals in a causal structure.

This temporarily causal model structure is a critical first step in the development of real-time ripple-replay detection and analysis methods. Multiple research groups have developed real-time methods for SWRs detection from LFP data and decoding from hippocampal spiking data separately in both hardware and software, but to our knowledge, no one has combined information from LFP rhythms and nonlocal spike patterns to detect ripple-replay events in a real-time manner. Gridchyn et al. developed a real-time decoding algorithm based on spike data to help detect high synchrony events in a closed-loop system using 24 kHz recordings \cite{b62}. A real-time spatial decoding algorithm using clusterless spikes was developed by Coulter et al. \cite{b87}. Siegle et al. developed a real-time spike detection method and applied it to raw signals recorded at a sampling rate of 30 kHz \cite{b51}. Xu et al. designed a real-time spike sorting and clustering algorithm and implemented it in a 16-channel processor \cite{b53}. Similarly, real-time collection and preprocessing of LFP signals has been implemented in many closed-loop experiments \cite{b51}, \cite{b52}. Oliva et al. developed real-time SWRs detection methods from single channel data from CA2 recorded at a sampling frequency of 20 kHz \cite{b57}. Stark et al. used two running windows to detect ripples in a closed-loop experiment \cite{b61}. Roux et al. used closed-loop SWRs detection at goal locations to trigger optogenetic silencing of a subset of CA1 pyramidal neurons using 20 kHz signals \cite{b88}. Gillespie et al. designed an online algorithm to detect SWRs from LFP data sampled at 30 kHz \cite{b52}. Dutta et al. analyzed an open source, closed-loop, real-time system for hippocampal sharp wave ripple (SWR) disruption with neural signal up to 30 kHz \cite{b92}. These studies demonstrate the ability to preprocess and extract information from LFP signals and spiking data in real-time.

The remainder of this paper is organized as follows. In Section II, we define the fundamental structure of the state-space model, and derive the algorithms for computing the filter and smoother probability distributions. In Section III, we present one simulation study in subsection A, and a real data analysis example in subsection B, to illustrate the application of these methods to data and highlight their estimation properties. In Section IV we discuss limitations and extensions for these methods.

\section{Methods
}

\subsection{Defining the State-Space Model}
Assume that the data are observed over a set of discrete time points $t_0, t_1, t_2, \ldots, t_k, \ldots, t_K$. We define a binary replay state process $I_k$, called the ripple-replay state, as follows:
\begin{equation}
    I_{k}=\left\{\begin{array}{ll}1, & \text { if a replay event is occurring at time } t_{k}; \\ 0, & \text { if a replay event is not occurring at time } t_{k};\end{array}\right.
\end{equation}
This is a latent state process, so we cannot directly observe the value of $I_k$. Instead, we will define another set of latent processes that link this replay state to the observation processes and then use the observed data to estimate the probability of being in a replay state at each instant. We assume the probability of being in a replay state at time $t_k$ only depends on the state at the previous time $t_{k-1}$. 

We link the ripple-replay state to the spiking activity from the population of place cells by defining a continuous state variable $x_k$, which captures the factors that influence spiking both in and out of the replay state. When out of the replay state, the neurons have place fields which fire as a function of the rat’s linearized position at time $t_k$. We let $o_k$ be the observed linearized position at time $t_k$. Therefore, whenever $I_k=0$, we set $x_k=o_k$. When $I_k=1$, we treat $x_k$ as an unobserved, latent variable. 

When the replay state changes from non-replay into replay, we allow $x_k$ to correspond to any location, not just the rat's current location. The distribution of possible initial locations for the replay event is assumed to be uniform over the entire environment. We assume that as long as the rat stays in the replay state, the trajectory of $x_k$ is continuous, and use a random walk to model transitions within this state. Mathematically, the dynamics of $x_k$ are defined by the following state transition model.
\begin{equation}
\label{eq0}
p\left(x_{k} \mid x_{k-1}, I_{k}, I_{k-1}\right)=\left\{\begin{array}{lc}\delta\left(o_{k}\right), & \text { if } I_{k}=0; \\ \mathcal{U}(a, b), & \text { if } I_{k}=1 \\ & \text { and } I_{k-1}=0; \\ \mathcal{N}\left(x_{k-1}, \sigma\right), & \text { if } I_{k}=1 \\ & \text { and } I_{k-1}=1;\end{array}\right.
\end{equation}

The spiking activity of each tetrode is modeled as a marked Poisson process \cite{b14}, which defines the probability distribution of observing a spike with particular waveform features as a function of $x_k$. When $I_k=0$, $x_k$ is the rat's actual position and so this model describes the place field properties of the entire population. When $I_k=1$, $x_k$ corresponds to an internal cognitive representation of position, and this model describes the pattern of reactivation of spiking during replay. In general, the marks can be any waveform features, up to the full waveform function. Here, we use the maximum amplitude of the spike on each of the 4 electrodes comprising the tetrode to define a 4-dimensional mark. The data comprises the spike time and the value of this mark for each observed spike. The marked point process model for the neural population recorded by tetrode $i$ is defined by a joint mark-intensity function $\lambda_{i}\left(t, \vec{m}\right)$, which expresses the probability of observing a spike with features in a neighborhood of $\vec{m}$ in an interval around $t$ \cite{b35}. We construct a joint mark-intensity model by expressing $\lambda_{i}\left(t, \vec{m}\right)$ as a function of $x_k$. 

If the observed spiking data from tetrode $i$ in the interval $[t_{k-1},t_k)$ includes $\Delta N_k^i$ spikes with marks $(\vec{m}_{k,1}^i, \vec{m}_{k,2}^i, \cdots \vec{m}_{k,\Delta N_k^i}^i)$, and we assume independence between tetrodes, the likelihood of the observed spiking is expressed as:
\begin{equation}
\label{likelihood}
L(x_{k}) \propto  \prod_{i=1}^{E} \prod_{j=1}^{\Delta N_{k}^{i}}\lambda_{i}\left(t_{k}, \vec{m}_{k, j}^{i} \right) \exp \left[-\Lambda_{i}\left(t_{k} \right) \Delta_{k}\right]
\end{equation}
where $E$ is the number of tetrodes, $\Delta_{k}=t_k-t_{k-1}$, and $\Lambda_{i}(t_{k})=\int_\mathcal{M} \lambda_{i}\left(t_{k}, \vec{m} \right) d\vec{m}$ is the integral of the joint mark intensity over the entire mark space, $\mathcal{M}$, called the ground intensity \cite{b32}. 

We link the ripple-replay state to the LFP activity across multiple electrodes through an additional set of latent vectors $\vec{Z}_k^d$, each of which defines an oscillator that governs one or more of the recorded LFP channels. A small number of oscillators, $D$, are posited to capture most of the rhythmic structure in the LFPs, and each $d \in {1,...,D}$ indexes one of these oscillators. Each oscillator is described as a latent, 2-dimensional vector autoregressive process of order 1, with a state transition matrix that is a rotation matrix that rotates $\vec{Z}_k^d$ in the 2D plane based on the frequency of the $d^{th}$ oscillator. The replay and nonreplay states have distinct sets of oscillators representing the distinct rhythmic content during active exploration and replay. That is to say, in the $k^{\text{th}}$ time interval, the state equation for the $d^{th}$ oscillator, $\vec{Z}_{k}^{d}$, given state $I_k$, is
\begin{equation}
    \left(\begin{array}{l}Z_{k,1}^{(d)} \\ Z_{k,2}^{(d)}\end{array}\right)=\vec{A}_{I_{k}, d}\left(\begin{array}{c}Z_{k-1, 1}^{(d)} \\ Z_{k-1, 2}^{(d)}\end{array}\right)+\left(\begin{array}{c}v_{I_k, 1}^{(d)} \\ v_{I_k, 2}^{(d)}\end{array}\right)
\end{equation}
where
\begin{equation}
    \vec{A}_{I_k, d}=a_{I_{k}, d}\left(\begin{array}{cc}\cos \left(2 \pi f_{I_{k}, d} \Delta_{k}\right) & -\sin \left(2 \pi f_{I_{k}, d} \Delta_{k}\right) \\ \sin \left(2 \pi f_{I_{k}, d} \Delta_{k}\right) & \cos \left(2 \pi f_{I_{k}, d} \Delta_{k}\right)\end{array}\right)
\end{equation}
and $(v_{I_{k}, 1}^{(d)}, v_{I_{k}, 2}^{(d)})^{\prime} \sim N_{2}(\mathbf{0}, \sigma_{I_{k}, d}^{2} \bold{I}_{2})$.

The observed hippocampal LFPs, $\vec{Y}_{k}$, can be written as a $J$-dimensional vector. The component from the $j^{th}$ channel is defined by
\begin{equation}
    y_{k, j}=\sum_{d=1}^{D}\left(c_{d, j, 1}^{I_k} Z_{k, 1}^{(d)}+c_{d, j, 2}^{I_k} Z_{k, 2}^{(d)}\right)+w_{k, j, I_k}
\end{equation}
where $c_{d, j, i}^{I_k}$ defines the contribution of each component of the $d^{th}$ oscillator to the $j^{th}$ LFP channel, for $i \in \{1,2\}$, and $w_{k, j, I_k} \sim N(0,\tau_{I_k}^2)$. The relative magnitude of $c_{d, j, 1}^{I_k}$ vs. $c_{d, j, 2}^{I_k}$ controls the phase of the $d^{th}$ rhythmic component expressed at the $j^{th}$ channel. To preserve identifiability we define $c_{d, 1, 1}^{I_k} = 1$ and $c_{d, 1, 2}^{I_k}=0$. 

\begin{figure}
\centerline{\includegraphics[width=8.8cm, height=7.8cm]{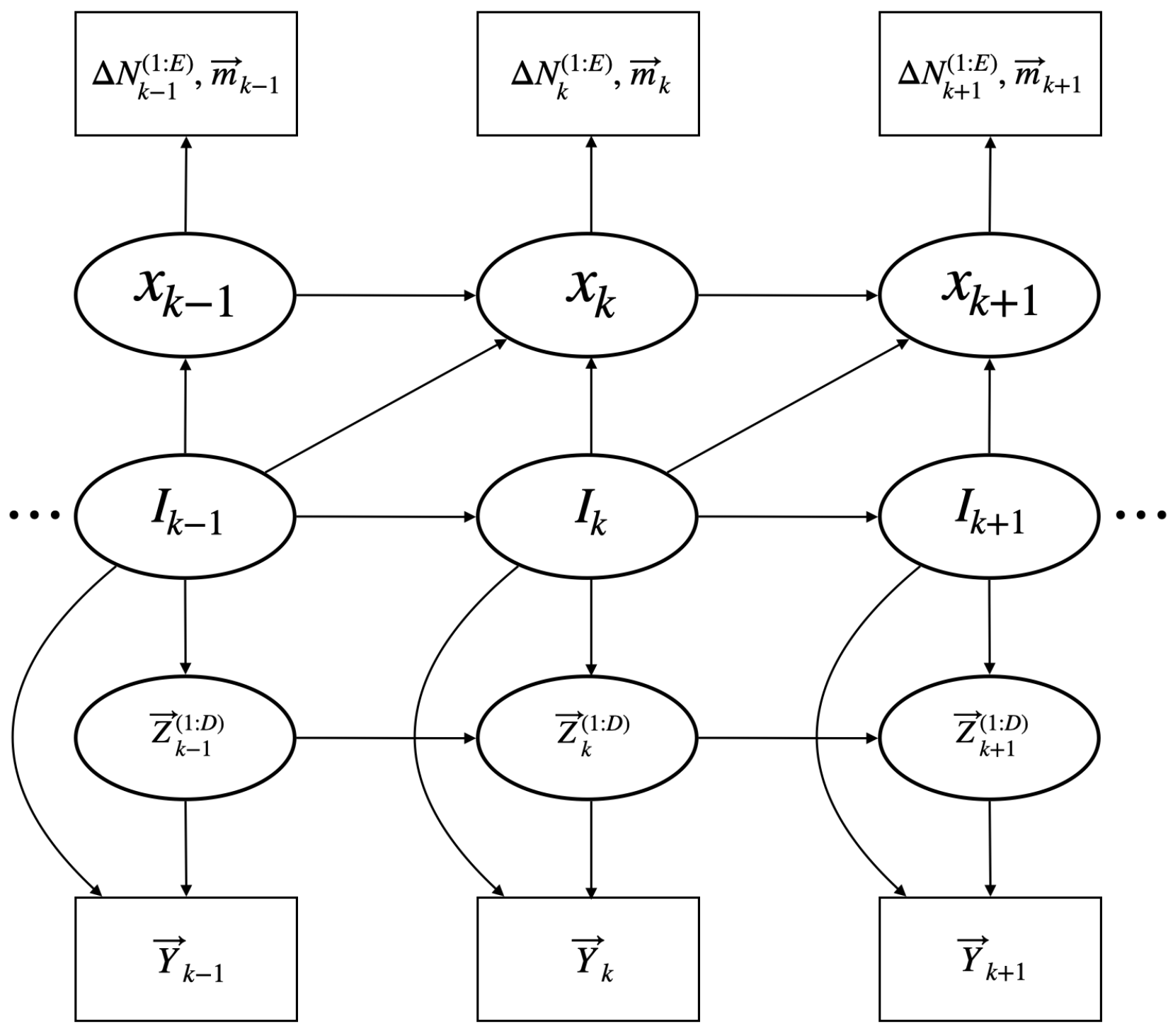}}
\caption{Graphical model representation. $I_k$, $x_k$ and $\vec{Z}_k^{(1:D)}$ are the latent variables. $\Delta N_{k}^{(1: E)}, \vec{m}_{k}$ and $\vec{Y}_k$ are the observations. Arrows indicate statistical dependence.}
\label{fig1}
\end{figure}

Fig. \ref{fig1} shows a graphical representation of the model structure. The latent variables are represented as ovals and the observations are represented as rectangles. Each column represents all of the variables at a single time point. The arrows depict the dependencies of each variable on the other variables at the current and next time step. The current semi-latent variable $x_k$ can be determined by previous value $x_{k-1}$, the previous replay state $I_{k-1}$ and the current state $I_k$. The current oscillator vector $\vec{Z}_k^{(1:D)}$ is determined by its previous value $\vec{Z}_{k-1}^{(1:D)}$ and the current replay state $I_k$. The set of spikes and waveforms during the current time step,  $\Delta N_{k}^{(1:E)}$ and $\vec{m}_k$, is related to $x_k$, which itself depends on the latent replay state $I_k$. The LFP observation $\vec{Y}_k$ is determined by the current replay state $I_k$ and the latent oscillators $\vec{Z}_{k}^{(1:D)}$. 

We note that this model structure assumes that the ripple-replay state, $I_k$, begins to influence the continuous states $x_k$ and $\vec{Z}_{k}^{(1:D)}$, and therefore the spike and LFP data across all channels at the same time. However, in real data, nonlocal spiking and ripple frequency activity often begins at different times for the same event. Below, we show that the decoder is able to capture information from multiple signals even when they start at different time points. Alternately, we could extend this model to explicitly include different time lags for different oscillators and LFP channels.

\subsection{Filter Solution}

In this section, we derive the specific solution for the probability of the replay state at each time step given the observed spiking and LFP data up to the current time, based on the state-space model discussed above. We derive the joint probability distribution for all latent variables, given all of the observed signals up to the current time i.e. $p(I_{k}, x_{k}, \vec{Z}_{k}^{(1:D)} \mid \vec{Y}_{k}, \Delta N_{k}^{(1: E)}, \vec{m}_{k}, H_{k})$, where $H_k$ is the history of the observed data up to, but not including $t_k$. This is called the filter distribution. This allows us to compute the marginal probability that a replay state is occurring at each instant as well as the marginal distributions of the latent variables representing the spatial content of each replay event and the small number of oscillators driving the LFPs.
	
First, using Bayes's rule, we express the filter distribution as the product of the likelihoods of the spiking and LFP data over the last time step with $p\left(I_{k}, x_{k}, \vec{Z}_{k}^{(1:D)} \mid H_{k} \right)$, the one-step prediction distribution of the states given all but the most recent observations.
\begin{equation}
\begin{aligned} \label{eq1}
& p\left(I_{k}, x_{k},\vec{Z}_{k}^{(1:D)} \mid \vec{Y}_{k}, \Delta N_{k}^{(1: E)}, \vec{m}_k,  H_{k}\right) \propto \\ & p\left(\vec{Y}_{k} \mid I_{k}, x_{k}, \vec{Z}_{k}^{(1:D)},\Delta N_{k}^{(1: E)}, \vec{m}_k,H_{k}\right) \\  & \times p\left(\Delta N_{k}^{(1: E)}, \vec{m}_k \mid I_{k}, x_{k}, \vec{Z}_{k}^{(1: D)}, H_{k}\right) \\  & \times p\left(I_{k}, x_{k}, \vec{Z}_{k}^{(1:D)} \mid H_{k} \right)
\end{aligned}
	\end{equation}
The first two terms on the right-hand side of (\ref{eq1}) are the likelihoods from each of the observation models, LFPs and population spiking activity. Combined with our assumptions in the previous section, these terms can be simplified.
	\begin{equation}
 \begin{aligned} \label{eq2}
	  &  p\left(I_{k}, x_{k}, \vec{Z}_{k}^{(1:D)} \mid \vec{Y}_{k}, \Delta N_{k}^{(1: E)}, \vec{m}_{k}, H_{k}\right) \propto \\ & p\left(\vec{Y}_{k} \mid I_{k}, \vec{Z}_{k}^{(1:D)} \right) p\left(\Delta N_{k}^{(1: E)}, \vec{m}_k \mid x_{k} \right) \\ & \times p\left(I_{k}, x_{k}, \vec{Z}_{k}^{(1:D)} \mid H_{k}\right)  
      \end{aligned}
	\end{equation}	
We use the Chapman-Kolmogorov equation to compute the one-step prediction distribution as a function of the filter distribution from the previous time step.
 \begin{equation}
 \begin{aligned} \label{eq3}
 & p\left(I_{k}, x_{k}, \vec{Z}_{k}^{(1:D)} \mid H_{k}\right)= \\ & \sum_{I_{k-1}} \int_{x_{k-1}} \int_{\vec{Z}_{k-1}^{(1:D)}} p\left(\vec{Z}_{k}^{(1:D)} \mid \vec{Z}_{k-1}^{(1:D)}, I_k\right) \\ & \times p\left(x_{k} \mid x_{k-1}, I_{k}, I_{k-1}\right) Pr\left(I_{k} \mid I_{k-1}\right) \\ & \times p\left(I_{k-1}, x_{k-1}, \vec{Z}_{k-1}^{(1:D)} \mid \vec{Y}_{k-1}, \Delta N_{k-1}^{(1: E)}, \vec{m}_{k-1}, H_{k-1}\right) \\ & d \vec{Z}_{k-1}^{(1:D)} d x_{k-1}
 \end{aligned}
	\end{equation}
The first term on the right hand side of (\ref{eq3}) is the transition density for the latent oscillator $\vec{Z}_{k}^{(1: D)}$, the second term is the transition density for semi-latent variable $x_k$ and the third term is the replay state transition probability for $I_k$. The last term is the filter distribution of the replay state, semi-latent variable and latent LFP oscillator from the previous time step. The sum and integrals serve to marginalize out the latent variables at the previous time step.

Combining (\ref{eq2}) and (\ref{eq3}), we obtain an iterative expression for the filter distribution.
  \begin{equation} \label{eq4}
  \begin{aligned}
 & p\left(I_{k}, x_{k},\vec{Z}_{k}^{(1:D)} \mid \vec{Y}_{k}, \Delta N_{k}^{(1: E)}, \vec{m}_{k},  H_{k}\right) \propto \\ & p\left(\vec{Y}_{k} \mid I_{k}, \vec{Z}_{k}^{(1:D)}\right) p\left(\Delta N_{k}^{(1: E)}, \vec{m}_{k} \mid x_{k}\right)  \\ & \times \sum_{I_{k-1}} \int_{x_{k-1}} \int_{\vec{Z}_{k-1}^{(1:D)}} Pr\left(I_{k} \mid I_{k-1}\right) \\ & \times p\left(\vec{Z}_{k}^{(1:D)} \mid I_k, \vec{Z}_{k-1}^{(1:D)} \right) p\left(x_{k} \mid x_{k-1}, I_{k}, I_{k-1}\right) \\ & \times p\left(I_{k-1}, x_{k-1}, \vec{Z}_{k-1}^{(1:D)} \mid \vec{Y}_{k-1}, \Delta N_{k-1}^{(1: E)}, \vec{m}_{k-1},  H_{k-1}\right) \\ & d \vec{Z}_{k-1}^{(1:D)} d x_{k-1}
 \end{aligned}
  \end{equation}

  The last term on the right side of (\ref{eq4}) is the filter distribution from the previous time step, which is multiplied by the state transition for each latent state and integrated and summed over the previous latent variables to produce the one-step prediction distribution. This distribution is then multiplied by the likelihood of the spiking and LFP data at the current time step, based on the observation model, to compute the filter distribution at the current time step. We select an initial distribution for the states at time $t_0$, and iterate through (\ref{eq4}) to compute the filter distribution at all time steps.

\subsection{Smoother Solution}

In the previous subsection, we derived the filter  distribution for all three latent variables, $I_k$, $x_k$ and $\vec{Z}_k^{(1:D)}$, which is based on the data until time $t_k$. We will now derive the partial smoother distribution for two of the latent processes, $I_k$ and $x_k$, as a function of all of the data through the last time step, $t_K$. That is to say, we derive an expression for the conditional distribution $p(I_{k}, x_{k} \mid \vec{Y}_{K}, \Delta N_{K}^{(1: E)}, \vec{m}_{K},  H_{K})$. Similar to the filter distribution, the smoother distribution can also be computed with an iterative formula that, in this instance, steps backward, using the smoother distribution from one step in the future to compute the distribution at the current time.

Analogous to (\ref{eq3}), we obtain the one-step prediction for $I_{k+1}$ and $x_{k+1}$ based on the observations up to $t_k$.
\begin{equation}
\begin{aligned} \label{eq5}
& p\left(I_{k+1}, x_{k+1} \mid \vec{Y}_{k}, \Delta N_{k}^{(1: E)}, \vec{m}_{k},  H_{k}\right)= \\  & \sum_{I_{k}} \int_{x_{k}} p\left(x_{k+1} \mid x_{k}, I_{k}, I_{k+1}\right) Pr\left(I_{k+1} \mid I_{k}\right) \\ & \times p\left(I_{k}, x_{k} \mid \vec{Y}_{k}, \Delta N_{k}^{(1: E)}, \vec{m}_{k},  H_{k}\right) d x_{k}
\end{aligned}
\end{equation}

Then, starting from the last estimate of the filter probability distribution $p(I_{K}, x_{K} \mid \vec{Y}_{K}, \Delta N_{K}^{(1: E)}, \vec{m}_{K}, H_{K})$, using both the filter distribution computed by iterating (\ref{eq4}) forward through all time steps, and the one-step prediction from (\ref{eq5}), we iterate the smoother distribution by stepping backwards as follows:

\begin{equation}
\begin{aligned} \label{eq6}
   & p\left(I_{k}, x_{k} \mid \vec{Y}_{K}, \Delta N_{K}^{(1: E)}, \vec{m}_{K}, H_{K}\right)  = \\ & p\left(I_{k}, x_{k} \mid \vec{Y}_{k}, \Delta N_{k}^{(1: E)}, \vec{m}_{k}, H_{k}\right)  \\ & \times \sum_{I_{k+1}} \int_{x_{k+1}} \frac{p\left(x_{k+1} \mid x_{k}, I_{k+1}, I_{k}\right) Pr\left(I_{k+1} \mid I_{k}\right)}{p\left(x_{k+1}, I_{k+1} \mid \vec{Y}_{k}, \Delta N_{k}^{(1: E)}, \vec{m}_{k}, H_{k}\right)} \\ & \times p\left(x_{k+1}, I_{k+1} \mid \vec{Y}_{K}, \Delta N_{K}^{(1: E)}, \vec{m}_{K}, H_K \right) d x_{k+1}
    \end{aligned}
\end{equation}

\subsection{Implementation Details}

The latent variables are multi-dimensional, and the integrals in (\ref{eq4}) and (\ref{eq6}) do not have analytic solutions. Therefore, we employ numerical integration methods, specifically sequential importance sampling to approximate the filter distribution. In the subsequent examples, we used the mixture Kalman filter to estimate the filter distribution sequentially \cite{b36}. This method provides sufficient accuracy with a small number of particles for these examples. The results below use $50$ particles, and we obtain similar results when the number of particles is increased. In order to deal with degeneracy, when the coefficient of variance of the particle weight is more than $1e^{-4}$, we resample each particle. For the smoother distribution, since $x_k$ is one-dimensional and $I_k$ is discrete, we computed a Riemann sum approximation with $\Delta x = 5$ cm in simulation and $\Delta x = 10$ cm in real data analysis to estimate the solution.

We made the following assumptions to define the initial condition for the states, $p(I_0, x_0, \vec{Z}_0^{(1:D)})$. We assumed that the binary state was initially to be in a non-replay state, $\operatorname{Pr}(I_{0}=0)=1$. By definition, when $I_0=0$, the distribution of $x_0$ is a delta function at the rat’s position. We initialized each oscillator with an independent zero-mean two-dimensional Gaussian distribution with covariance matrix $\sigma_{0, d}^{2} \bold{I}_{2}$.

\section{Results}

\subsection{Simulation Study}

We simulated data from a rat performing a 10-minute spatial navigation task on a linear track, generating the movement, LFP, and spiking data as well as transitions between the ripple-replay state. We first sampled the latent replay state $I_k$ using the transition probabilities $\operatorname{Pr}\left(I_{k+1}=1 \mid I_{k}=0\right)=0.0003$ and $\operatorname{Pr}\left(I_{k+1}=1 \mid I_{k}=1\right)=0.99$. We simulated the movement trajectory $o_k$ starting from the position $o_0 = 90$ cm using a random walk with standard deviation $0.5$ cm for each time step. The time step in our simulation was chosen to be $0.667$ ms. The path was bounded between $0$ cm and $187$ cm by fixing any movements that would have extended beyond these bounds at the boundary. We sampled the semi-latent state $x_k|I_k$ using ($\ref{eq0}$), where the range of the uniform distribution was from $0$ to $187$ cm and the standard deviation for the random walk was $0.5$ cm. 

Finally, two channels of LFP and $9$ tetrodes of spike data were generated along with waveform features for each spike. The simulated LFP data used $4$ pairs of oscillators with the parameters given in Tables \ref{t1} and \ref{t2}. When $I_k=0$, the LFP spectrum has peaks at about $10$ hz, $60$ hz and $80$ hz. When $I_k=1$, the spectral peaks are near $10$ hz, $80$ hz and $170$ hz. 

\begin{table}
\renewcommand{\arraystretch}{1}
\caption{Simulation Parameters When $I_k=0$}
\label{t1}
\centering
\setlength{\tabcolsep}{1mm}
\begin{tabular}{cc}
\hline
$\text{Parameters}$ & $I_k=0$ \\
\hline
 $\left(a_{I_k,1}, a_{I_k,2}, a_{I_k,3}, a_{I_k,4}\right)$ & $\left(0.98, 0.98, 0.98, 0.98\right)$  \\
\hline
$\left(f_{I_k,1}, f_{I_k,2}, f_{I_k,3}, f_{I_k,4} \right)$ & $\left(7, 9, 58, 82\right)$  \\

\hline
$\left(\sigma_{I_k,1}^{2}, \sigma_{I_k,2}^{2}, \sigma_{I_k,3}^{2}, \sigma_{I_k,4}^{2}\right)$ & $\left(25, 25, 25, 25\right)$ \\
\hline
$\left(c_{1,2,1}^{I_k}, c_{1,2,2}^{I_k}, \cdots, c_{4,2,1}^{I_k}, c_{4,2,2}^{I_k}\right)$ & $\left(1, -0.5, -0.5, 1, 0.5, -1, -0.5, 1\right)$  \\
\hline
$\tau_{I_k}^{2}$ & $1$ \\
\hline 
\end{tabular}
\end{table}

\begin{table}
\renewcommand{\arraystretch}{1}
\caption{Simulation Parameters When $I_k=1$}
\label{t2}
\centering
\setlength{\tabcolsep}{1mm}
\begin{tabular}{cc}
\hline
$\text{Parameters}$ & \bfseries $I_k=1$ \\
\hline
 $\left(a_{I_k,1}, a_{I_k,2}, a_{I_k,3}, a_{I_k,4}\right)$ & $\left(0.98, 0.98, 0.98, 0.98\right)$ \\
\hline
$\left(f_{I_k,1}, f_{I_k,2}, f_{I_k,3}, f_{I_k,4} \right)$ & $\left(5 , 14  ,80 , 168\right)$  \\

\hline
$\left(\sigma_{I_k,1}^{2}, \sigma_{I_k,2}^{2}, \sigma_{I_k,3}^{2}, \sigma_{I_k,4}^{2}\right)$ & $\left(25, 25, 25, 25\right)$ \\
\hline
$\left(c_{1,2,1}^{I_k}, c_{1,2,2}^{I_k}, \cdots, c_{4,2,1}^{I_k}, c_{4,2,2}^{I_k}\right)$ & $\left(1, -0.5, -0.5, 1, 0.5, -1, -0.5, 1\right)$  \\
\hline
$\tau_{I_k}^{2}$ & $1$ \\
\hline 
\end{tabular}
\end{table}

We simulated the spike times and waveforms for the first two tetrodes using a joint mark intensity that included two modes corresponding to two place cells:
\begin{equation}
\footnotesize
\lambda_{i}\left(t_{k}, \vec{m}\right)\propto \sum_{j=1}^2 \exp \left\{-\frac{1}{2}\left(\begin{array}{c}x_k-\mu_{x,j}^{i} \\ \vec{m}-\mu_{\vec{m},j}^{i}\end{array}\right)^{\prime}\Sigma^{-1}\left(\begin{array}{c}x_k-\mu_{x,j}^{i} \\ \vec{m}-\mu_{\vec{m},j}^{i}\end{array}\right)\right\}
\end{equation}
where $i=1,2$, $\mu_{x,1}^1=10$, $\mu_{x,2}^1=170$, $\mu_{x,1}^2=50$, $\mu_{x,2}^2=150$, $\mu_{\vec{m},1}^{1}=(370,370,370,370)^{\prime}$, $\mu_{\vec{m},2}^{1}=(230,230,230,230)^{\prime}$, $\mu_{\vec{m},1}^{2}=(350,350,350,350)^{\prime}$ and $\mu_{\vec{m},2}^{2}=(250,250,250,250)^{\prime}$. The mode of the ground intensity function is set to be $400$ Hz. $\Sigma$ is the covariance matrix with the following form:
{\small
\renewcommand{\arraycolsep}{1pt} 
\begin{equation}
\Sigma =
\begin{pmatrix}
\sigma_x & 0 & 0 & 0 & 0 \\
0 & \sigma_m & 0 & 0 & 0 \\
0 & 0 & \sigma_m & 0 & 0 \\
0 & 0 & 0 & \sigma_m & 0 \\
0 & 0 & 0 & 0 & \sigma_m
\end{pmatrix}
\begin{pmatrix}
1 & 0.6 & 0.8 & 0.7 & 0.9 \\
0.6 & 1 & 0.7 & 0.8 & 0.6 \\
0.8 & 0.7 & 1 & 0.6 & 0.8 \\
0.7 & 0.8 & 0.6 & 1 & 0.7 \\
0.9 & 0.6 & 0.8 & 0.7 & 1
\end{pmatrix}
\begin{pmatrix}
\sigma_x & 0 & 0 & 0 & 0 \\
0 & \sigma_m & 0 & 0 & 0 \\
0 & 0 & \sigma_m & 0 & 0 \\
0 & 0 & 0 & \sigma_m & 0 \\
0 & 0 & 0 & 0 & \sigma_m
\end{pmatrix}
\end{equation}
}
We chose this form of $\Sigma$ with non-zero off-diagonal elements to be consistent with real data for which the spike amplitudes from each channel are highly correlated. For this simple simulation scenario, we choose $\sigma_x=10$ and $\sigma_m=10$. The remaining tetrodes had a joint mark intensity with a single place cell:
\begin{equation}
\footnotesize
\lambda_{i}\left(t_{k}, \vec{m}\right)\propto \exp \left\{-\frac{1}{2}\left(\begin{array}{c}x_k-\mu_{x}^{i} \\ \vec{m}-\mu_{\vec{m}}^{i}\end{array}\right)^{\prime}\Sigma^{-1}\left(\begin{array}{c}x_k-\mu_{x}^{i} \\ \vec{m}-\mu_{\vec{m}}^{i}\end{array}\right)\right\}
\end{equation}
where $\mu_x^{i}=20i-40$, $\mu_{\vec{m}}^{i}=(20i+160,20i+160,20i+160,20i+160)^{\prime}, i=3,4,\cdots 9$. The mode of ground intensity function as well as $\Sigma$ are the same as the first two tetrodes.

\begin{figure}
\centerline{\includegraphics[width=8.8cm, height=8.4cm]{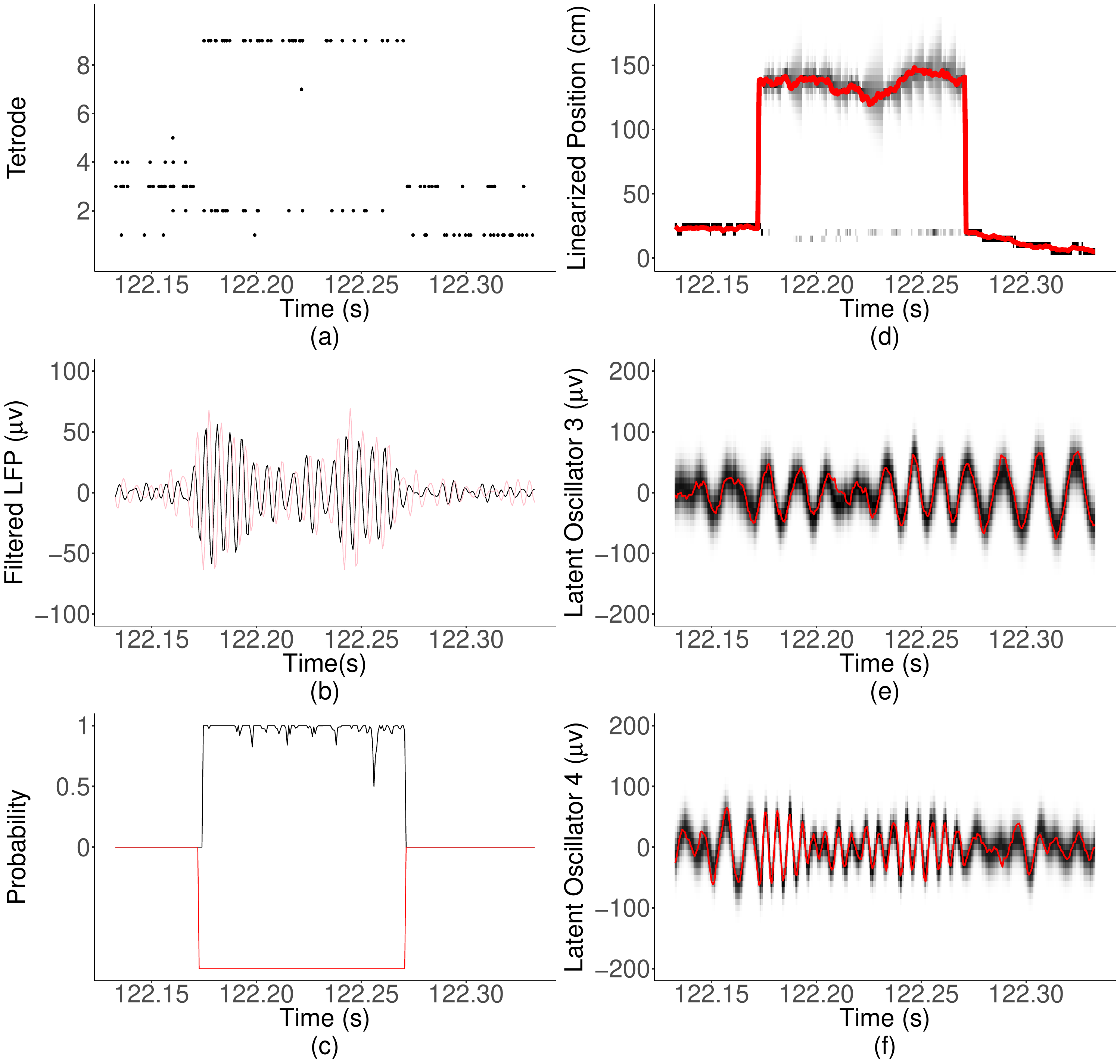}}
\caption{Decoding example for simulated data. (a) Simulated hippocampal spike data from 9 tetrodes. (b) LFPs from two channels (one black, the other red), filtered between $150$ Hz and $250$ Hz. (c) Filter probability of a replay event at each time step. The red line is the actual latent replay state. (d) Heat plot of the filter probability distribution for the semi-latent continuous variable, $x_k$. The red line is the actual trajectory of the semi-latent variable. (e) and (f) Heat plot of filter distribution for the first component from $3^{\text{rd}}$ and $4^{\text{th}}$ pairs of latent oscillators. The red line is the actual first component of each pair of latent oscillator.}
\label{fig2}
\end{figure}

We assumed all the true model parameters were known and decoded all of the latent variables ($I_k, x_k, Z_k^{(1:D)}$) simultaneously. An example segment of the decoding result is shown in Fig. \ref{fig2}. Fig. \ref{fig2}(a) shows the simulated hippocampal spike data from 9 tetrodes. There is an evident switch in the spiking pattern that occurs due to the occurrence of a ripple-replay event. Fig. \ref{fig2}(b) shows 2 channels of LFP data filtered between the ripple range of $150$ Hz to $250$ Hz. Fig. \ref{fig2}(c) shows the filter probability of a replay event at each time step. The red line is the actual latent replay state (inverted for visual clarity). When the actual state shifts to a ripple-replay event, the filter probability rapidly switches from a value near 0 to a value near 1, and persists at that high value throughout the replay event, after which it rapidly switches back. The filter probability does dip occasionally, when the spiking and LFP data become more ambiguous. Fig. \ref{fig2}(d) shows the value of the simulated semi-latent process in red, and a heat plot of its filter probability in grayscale. The center of the filter distribution rapidly jumps to, and then hovers around, the true value of $x_k$ when the switch to a ripple-replay event occurs.  Fig. \ref{fig2}(e) and (f) show the true values of the first component from the $3^{\text{rd}}$ and $4^{\text{th}}$ pairs of oscillators that generate the LFP signals in red, and the heat plots of the estimated filter probability distribution corresponding to each oscillator in grayscale. The ripple-replay period is associated with an increase in frequency these oscillators, which is accurately estimated by the filter. 

\begin{table}
\renewcommand{\arraystretch}{1}
\caption{Ripple-replay detection from state-space model}
\label{t3}
\centering
\setlength{\tabcolsep}{1mm}
\begin{tabular}{ccc}
\hline
\bfseries  & True Replay Event & No Replay Event\\
\hline
 Predicted Replay Event & 194 & 0 \\
\hline
No Predicted Replay Event & 4 & * \\
\hline
\end{tabular}
\end{table}

Table \ref{t3} shows the number of true positives, false positives, and false negatives for detecting ripple-replay events for this simulation analysis. We defined a detection whenever the filter probability of a replay event exceeded $0.99$ and stayed high for at least $15$ ms. Of the $198$ actual simulated ripple replay events, the filter was able to correctly identify $194$, and missed $4$. At all other times, no replay events occurred or were detected.

We compared the accuracy of our method with existing methods that use information exclusively from LFPs or exclusively from nonlocal spiking using this simulated data. The first method uses two running windows to detect SWRs using only past LFP data. It detects a ripple whenever the root-mean-square (RMS) of the band-pass filtered (150–250 Hz) signal from a short  window at the current time exceeds for at least $8$ ms three times RMS from a longer window extending back \cite{b61}. We applied this method to each LFP channel and computed the union of the detected ripple intervals. The second methods was a non-causal ripple detector that uses past and future LFP data \cite{b38}. This method identifies ripples as periods where the z-scored values of the population power trace exceeds the value $2$ for a minimum of $15$ ms. The ripple period is then extended forward and backward from the times of these detections for as long as the population power exceeds the mean.

We also compared our method to one that detects periods of nonlocal activity based on past hippocampal spiking patterns \cite{b89}. In this detection algorithm, burst and content detections are combined. The burst detection component responds to the crossing of a threshold $10$ Hz in the causal moving average of multi-unit activity (MUA) from any tetrode. The content detection component searches for sharp high-fidelity position estimates that are consistently located in nearby regions during last $10$ ms. The sharpness of the estimates is computed as the causal moving average of the locally integrated probability around the maximum-a-posteriori (MAP) estimate. The posterior position estimates come from the likelihood of observed spiking as described in (\ref{likelihood}) combined with a uniform prior. A replay event is defined if both conditions are met simultaneously.

\begin{figure}
\centerline{\includegraphics[width=9cm, height=4cm]{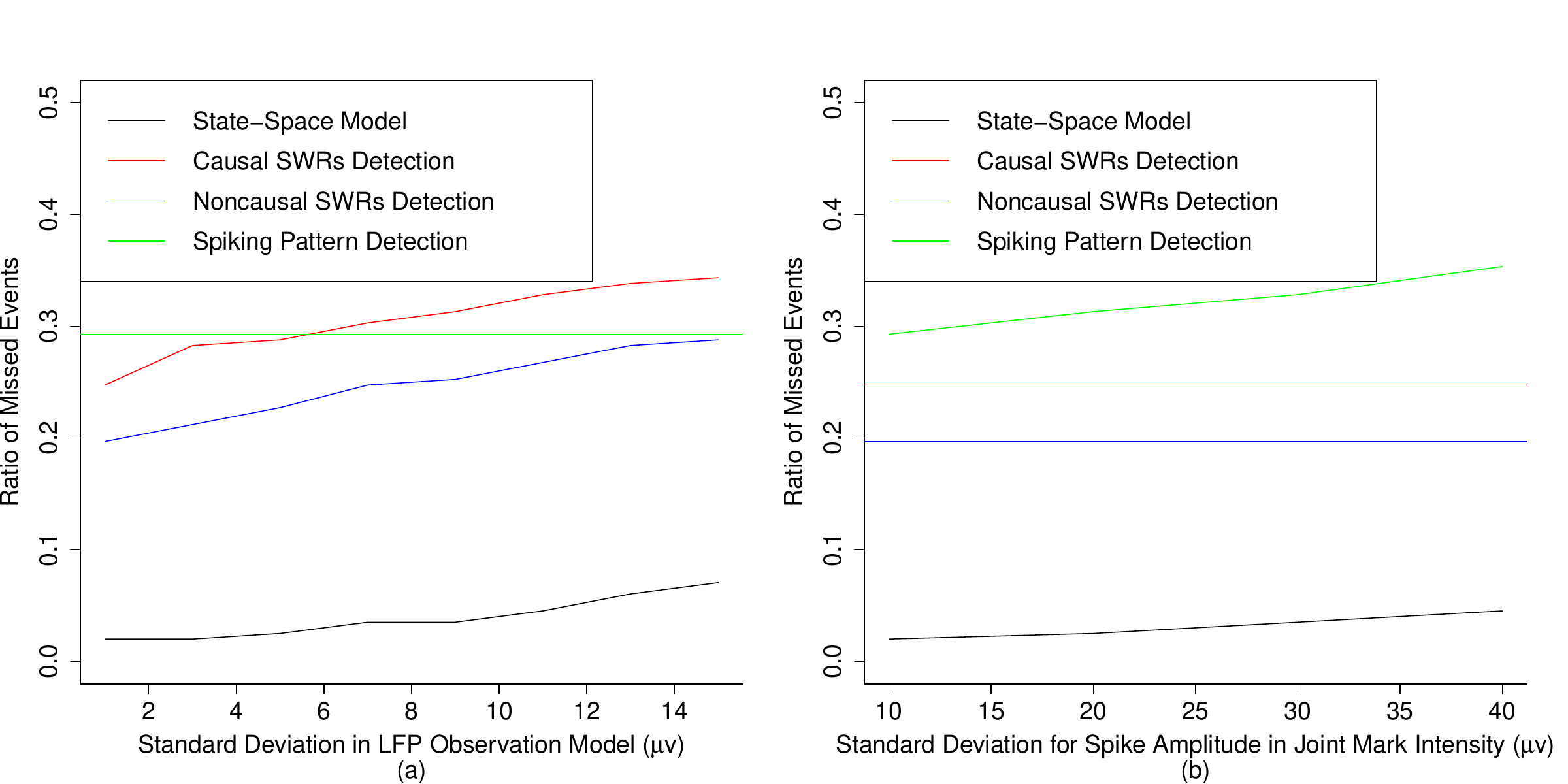}}
\caption{Missed detection ratio in different simulation scenarios. (a) Missed detection ratio for each method as a function of the noise in the LFP observation model, $\tau_{I_k}$. (b) Missed detection ratio as a function of the different joint mark intensity standard deviation $\sigma_m$. For each plot, the black line shows the missed detection ratio for our state-space model, the red line shows the missed detection ratio from the causal SWRs detection while the blue line shows results based on noncausal SWRs detection, and the green line shows the results from spiking pattern detection method. }
\label{fig8}
\end{figure}

Fig. \ref{fig8} shows the fraction of ripple replay events that are missed for each of the detection methods. Fig. \ref{fig8}(a) shows the missed detection ratio as a function of the LFP observation model standard deviation $\tau_{I_k}$. There is a clear gap between our methods, which uses information from both the LFP and spikes, and the other methods which use information from only one of these sources. As the LFP noise level increases, the three methods that use LFP information have increased numbers of missed detections, but this gap persists. Fig. \ref{fig8}(b) shows the missed detection ratio when the uncertainty in the standard deviation of marks in the joint mark intensity, $\sigma_m$, increases. As spikes become less informative, the two methods that use spike information have increased numbers of missed detections but the gap between these methods stays consistent. This simulation scenario was designed highlight the potential advantage of using information from both LFP and nonlocal spiking data.

\begin{figure}
\centerline{\includegraphics[width=8.8cm, height=7cm]{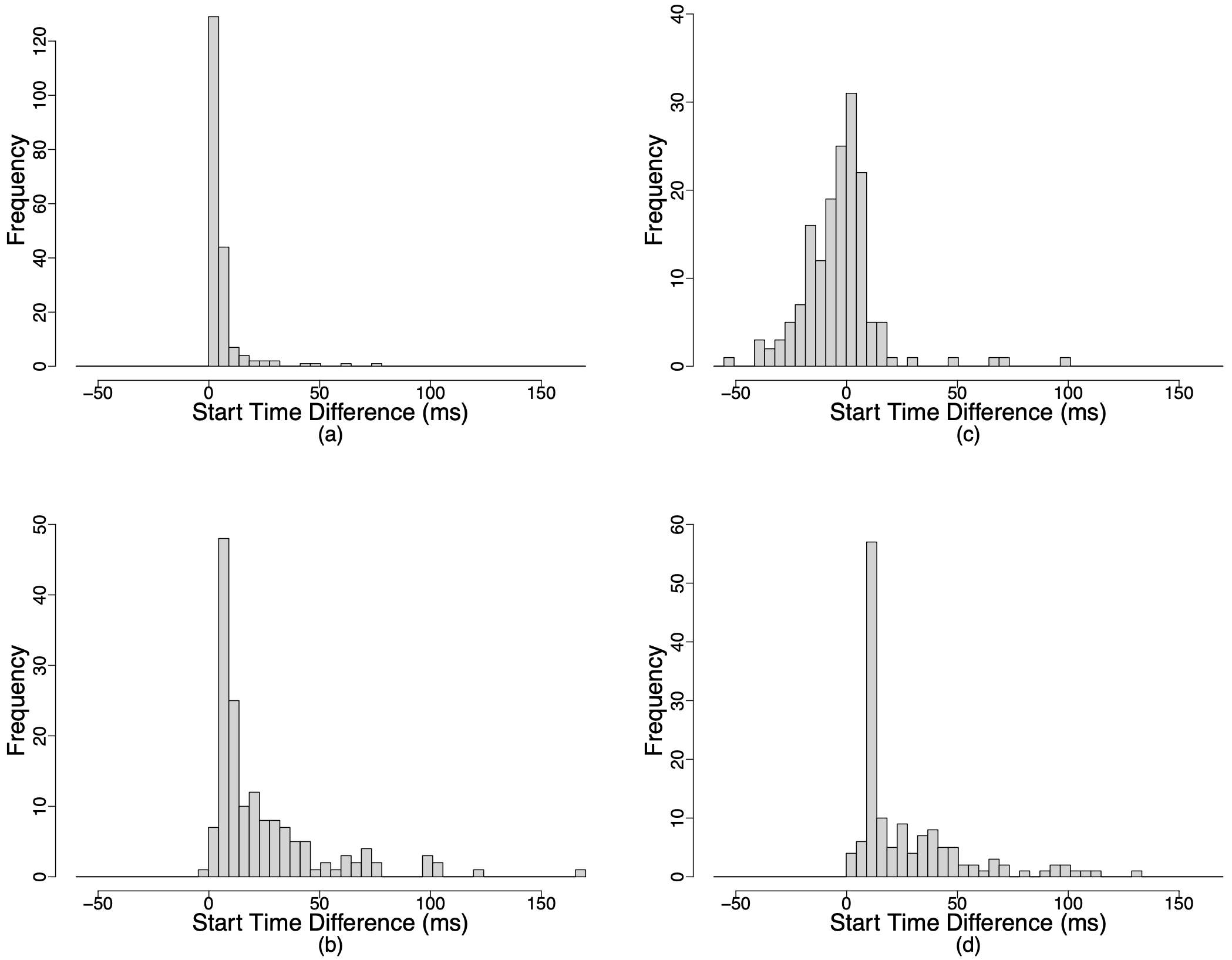}}
\caption{Histogram of detection time delay between true state and several detection methods. (a) Detection delay from our state-space model. Events whose decision state probability passes the threshold $0.99$ with duration of high probability more than $15$ ms are counted. (b) Detection delay from causal SWRs detection (c) Detection delay from noncausal SWRs detection. (d) Detection delay based on spiking pattern detection method.}
\label{fig3}
\end{figure}

We also compared the detection delay between our state-space model and the other three detection methods. Fig. \ref{fig3} shows a histogram of the difference in the start time between the estimated transitions to ripple-replay events. Fig. \ref{fig3}(a) shows the detection delay based on our state-space model. The estimated transitions were defined when the filter estimates exceeded a certainty level of $0.99$. Different choices for the certainty threshold (e.g. $0.9$ or $0.95$) lead to similar results. The decoder identifies replay events with strong confidence and minimal delay, using only past LFP and spiking information. Fig. \ref{fig3}(b) shows the delay from the detection method using only past LFP data. The detection delay is larger on average since more time was required to accumulate enough information from the past LFP signal alone. Fig. \ref{fig3}(c) shows the delay using past and future LFP data. This leads to detections that can precede the actual events. The variance of these detections is substantially larger than our state-space estimation method. Fig. \ref{fig3}(d) displays the delay from the detection method based only on past spiking activity. Again, the detection delays are larger on average, since some of the information from the LFP is missing. Notably, while the delays of other detection methods vary widely, our approach not only responds swiftly to ripple-replay events but also maintains a consistently narrow delay range, which demonstrates the robustness of this method. 

\begin{figure}
\centerline{\includegraphics[width=8.8cm, height=5.3cm]{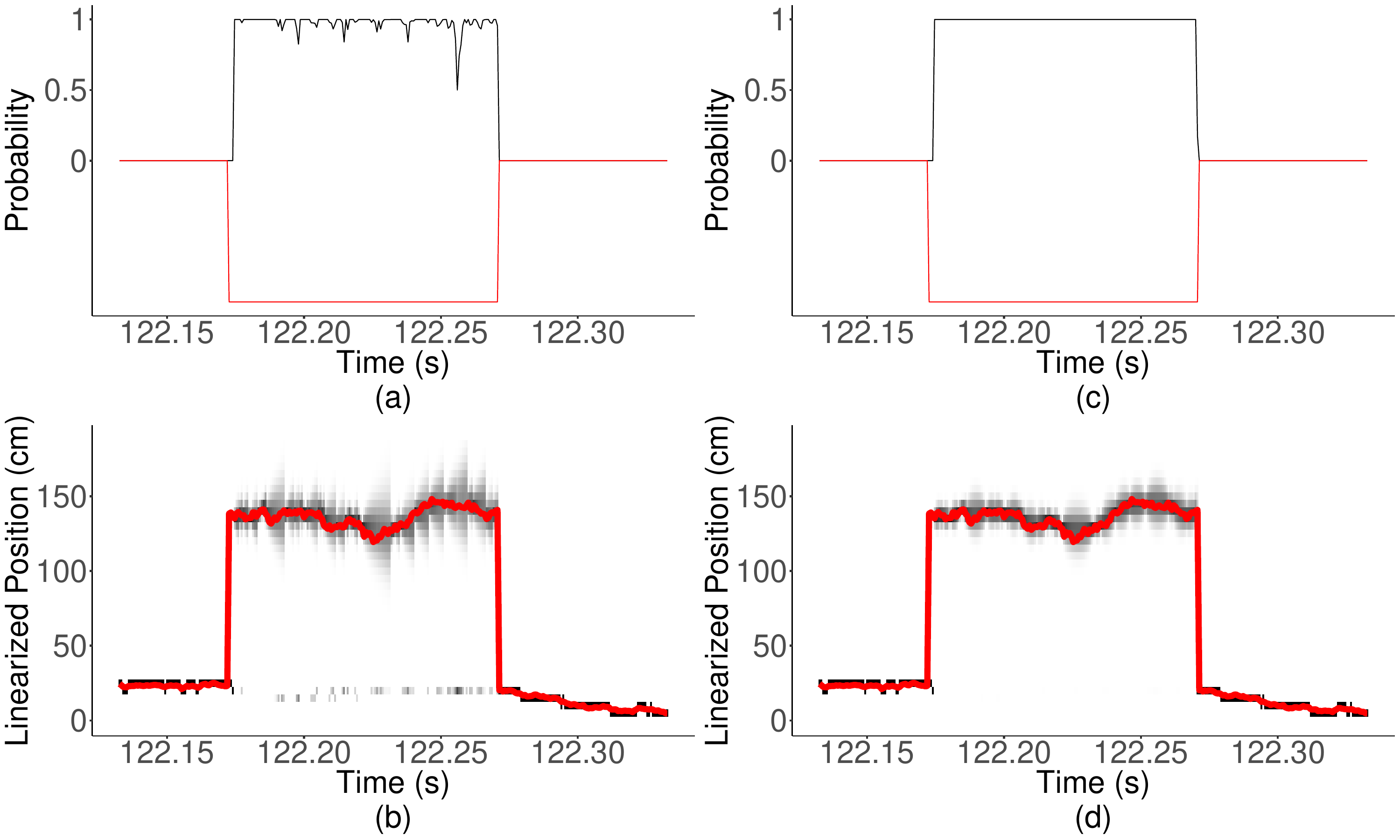}}
\caption{Comparison of filter and smoother decoding estimates for simulated data. (a) Filter probability of a replay event at each time step. The red line is the actual latent replay state. (b) Heat plot of filter probability distribution for semi-latent continuous variable. The red line is the actual semi-latent variable, and the shadow area is the posterior distribution. (c) Smoother probability of a replay event at the current time. (d) Heat plot of smoother probability distribution for semi-latent continuous variable. The red line is the actual semi-latent variable, and the shadow area is the posterior distribution.}
\label{fig4}
\end{figure}

The fact that the filter uses only past information from the LFP and spiking data makes it appropriate for real-time estimation. If instead we have access to past and future LFP and spike data, and would like to refine the filter estimates, we can decode using the smoother algorithm. We expect the smoother estimates to have smaller variability and better accuracy than the filter. A comparison between filter and smoother probability distribution for simulated data is shown in Fig. \ref{fig4}. Fig. \ref{fig4}(a) shows the filter probability distribution of a replay event at each time step. The red line is the actual latent replay state. Fig. \ref{fig4}(b) shows a heat plot of filter probability distribution for the state variable, $x_k$. The red line is the actual value of $x_k$, and the grayscale heatmap is the filter distribution. Fig. \ref{fig4}(c) shows the smoother probability distribution of a replay event at each time step. Compared with the filter, the smoother distribution does not dip down moment-to-moment, even when the spiking and LFP data become more ambiguous. Fig. \ref{fig4}(d) shows a heat plot of smoother probability distribution for $x_k$. As with the filter, the center of the smoother distribution rapidly jumps to the true value of $x_k$ when the ripple-replay event occurs. Compared to the filter, the smoother distribution is more accurate and more concentrated around the true value of $x_k$ with less variability. 

\subsection{Hippocampal Data Analysis Results}

 We analyzed publicly available hippocampal data from a single male Long Evans rat trained to perform a memory-guided spatial alternation task on a maze. The data, recorded in the laboratory of Dr. Anna Gillespie is available online at \url{https://dandiarchive.org/dandiset/000115/0.210914.1732}. We extracted two channels of LFPs and spiking data from $20$ tetrodes located in the CA1 region during a single $10$ minute trial of spatial navigation.

 The raw neural signal data were collected with a sampling rate of 30 kHz. Hippocampal LFP traces were generated for each tetrode by filtering the continuous signal from one channel of each tetrode between 1-300 Hz and then downsampled to 1500 Hz. In parallel, the continuous signal was referenced and filtered between 600-6000 Hz, and spike events were detected when the voltage exceeded 100 $\mu$V in any channel of a tetrode. The neural signal filtering process was implemented with the Butterworth filter, which used only past signal information.

Decoding was performed using a $\Delta_k = 0.667$ ms time step. For the state update model in (\ref{eq0}), we set the uniform distribution to the full range of the linearized position, $[240, 780]$ cm, and the random walk standard deviation to $10$ cm. The parameters for the Markov update model of the replay state were $\operatorname{Pr}(I_{k}=1 \mid I_{k-1}=0)=0.001$ and $\operatorname{Pr}(I_{k}=1 \mid I_{k-1}=1)=0.99$. We used a range of state transition probabilities to decode, and found that the results (eg. number of detected replay events, event duration, event start time, event end time etc.) were consistent across a large range of values. This demonstrates that our model is not overly sensitive to the choice of state transition probability. Finally, we chose a the transition probability values so that the expected duration of each ripple was about $67$ ms, consistent with previous findings \cite{b29}, \cite{b84}.

For the clusterless spike, we used a kernel based intensity model fit over the spiking data to estimate intensity function from a separate training set of the form: 
\begin{equation}
\label{lambda}
\lambda_{i}\left(t, \vec{m}\right) = \frac{\sum_{i=1}^N \mathcal{K}(x_t-x_{i},\vec{m}-\vec{m}_{i})}{\sum_{j=1}^K \mathcal{G}(x_t-x_j)}
\end{equation}
where $N$ is the total number of spikes in the training set, $\mathcal{K}(x_t-x_{i},\vec{m}-\vec{m}_{i})$ is a 5-dimensional Gaussian kernel function (1 dimension for space and 4 for each mark dimension), centered at each observed spike location and mark in the training set, and $\mathcal{G}(x_t-x_j)$ is a 1-dimensional Gaussian kernel function centered at each location visited at the $K$ discrete time points in the training set. We used a bandwidth of $10$ cm for the linearized position and $20$ $\mu$V for the dimensions of the spike amplitude.

For the LFP state and observation models, we applied an empirical Bayes method to estimate the model parameters \cite{b15}, \cite{b16}. For this analysis, we used $6$ pairs of LFP oscillators, which minimized the Akaike information criterion. We also performed a sensitivity analysis by varying the number of oscillators (ie. 4 to 8), and found minimal differences in the detections in terms of the number and duration of detected events.

\begin{figure}
\centerline{\includegraphics[width=8.8cm, height=8.4cm]{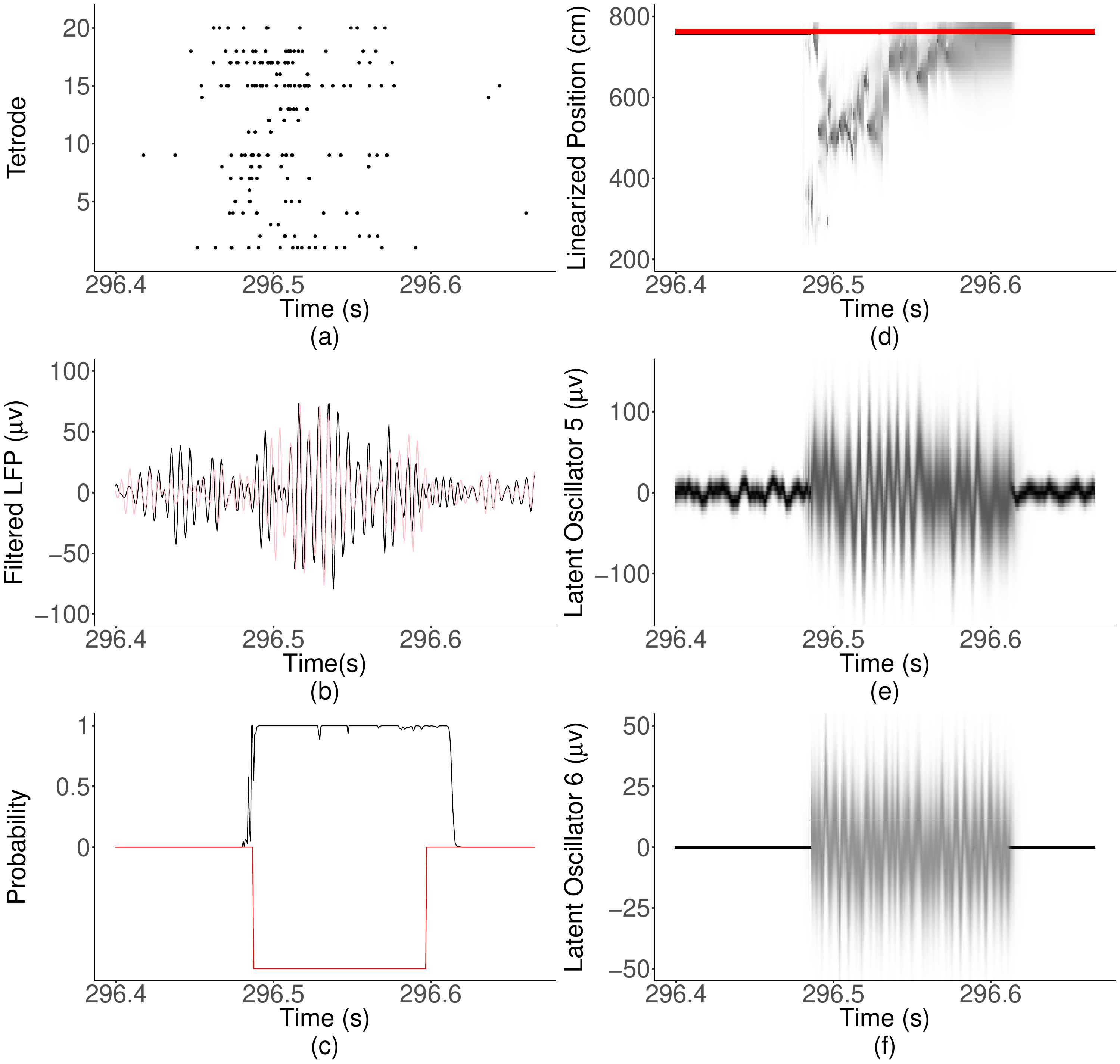}}
\caption{Decoding example for real data. (a) The raw hippocampal spike data from $20$ tetrodes. (b) LFPs from two channels (shown in red and black), filtered between 150-250 Hz. (c) Filter probability of a replay event at current time. The red line is the result from noncausal SWRs detection. (d) Heat plot of filter probability distribution for semi-latent continuous variable $x_k$. The red line is the observed linearized position, and the shadow area is the distribution. (e)-(f) Heat plot of filter probability distribution for the first component from $5^{\text{th}}$ and $6^{\text{th}}$ pairs of latent oscillators. The shadow area is the distribution.}
\label{fig5}
\end{figure}

Fig. \ref{fig5} shows an example decoding result from this dataset. Fig. \ref{fig5}(a) shows unclustered spike data from 20 tetrodes in hippocampus. Fig. \ref{fig5}(b) shows the LFP data over 2 channels, filtered between 150-250 Hz. Fig. \ref{fig5}(c) shows the estimated probability distribution of a replay event in black and a noncausal estimate of the ripple period based on the method in \cite{b38} in red. The filter probability switches from a value near 0 to a value near 1, and persists at that high value throughout the replay event with little fluctuation, which expresses certainty about the detection. There is little delay between the detection of the replay event by the causal filter the noncausal detector. In this example we can see the nonlocal spiking activity occur before an evident increase in LFP power in the ripple band. The increase in ripple power also does not occur identically across the two LFP channels. Even though there is not synchronization among the sources of neural information, our method was still able to detect ripple-replay events with high confidence and minimal delay. Fig. \ref{fig5}(d) shows a heat plot of filter probability distribution for the semi-latent continuous variable $x_k$. When the state changes from non-replay to replay, the representation could jump to a new location. Fig. \ref{fig5}(e) and (f) show heat plots of the filter probability distribution for the first component of each of the $5^{\text{th}}$ and $6^{\text{th}}$ pairs of latent oscillators. The ripple-replay period is associated with a significant change in the frequency and variance in the two sets of oscillators.

\begin{table}
\renewcommand{\arraystretch}{1}
\caption{Comparison Between Filter Probability and SWRs Detection}
\label{t4}
\centering
\setlength{\tabcolsep}{1mm}
\begin{tabular}{ccc}
\hline
\bfseries  &  Filter Probability: Yes &  Filter Probability: No\\
\hline
 SWR Detection: Yes & 214 & 106 \\
\hline
SWR Detection: No & 304 & * \\
\hline
\end{tabular}
\end{table}

While ground truth information about the occurrence of ripple-replay events was available in our simulated data, no such ground truth is available for this real data analysis. However, we can compare our detections to those obtained by the SWRs detector that used past and future LFP data to identify events for which both methods agree, and those that are detected by only one method. Table \ref{t4} shows the number of detected ripple-replay events reported by the two methods. The definition of a detection from filter probability is the same as the simulation study: whenever the probability of an event exceeded $0.99$ and kept high probability for at least $15$ ms. $214$ events were detected by both methods. $106$ events were only detected by the noncausal detection method. While $304$ were identified by our method that could not be detected by the noncausal method. There are a number of potential explanations for the large number of detections that were missed by the noncausal method. One possibility is that there are many periods where nonlocal spiking is present but no substantial high-frequency activity is observed in the LFPs. Another possibility is that many events include both changes in the population spiking and LFPs, but there is insufficient information in the LFPs alone to estimate these events with confidence; the state-space model is able to integrate information from both spikes and LFPs to make detections that the prior method misses. 

\begin{figure}
\centerline{\includegraphics[width=8.8cm, height=5cm]{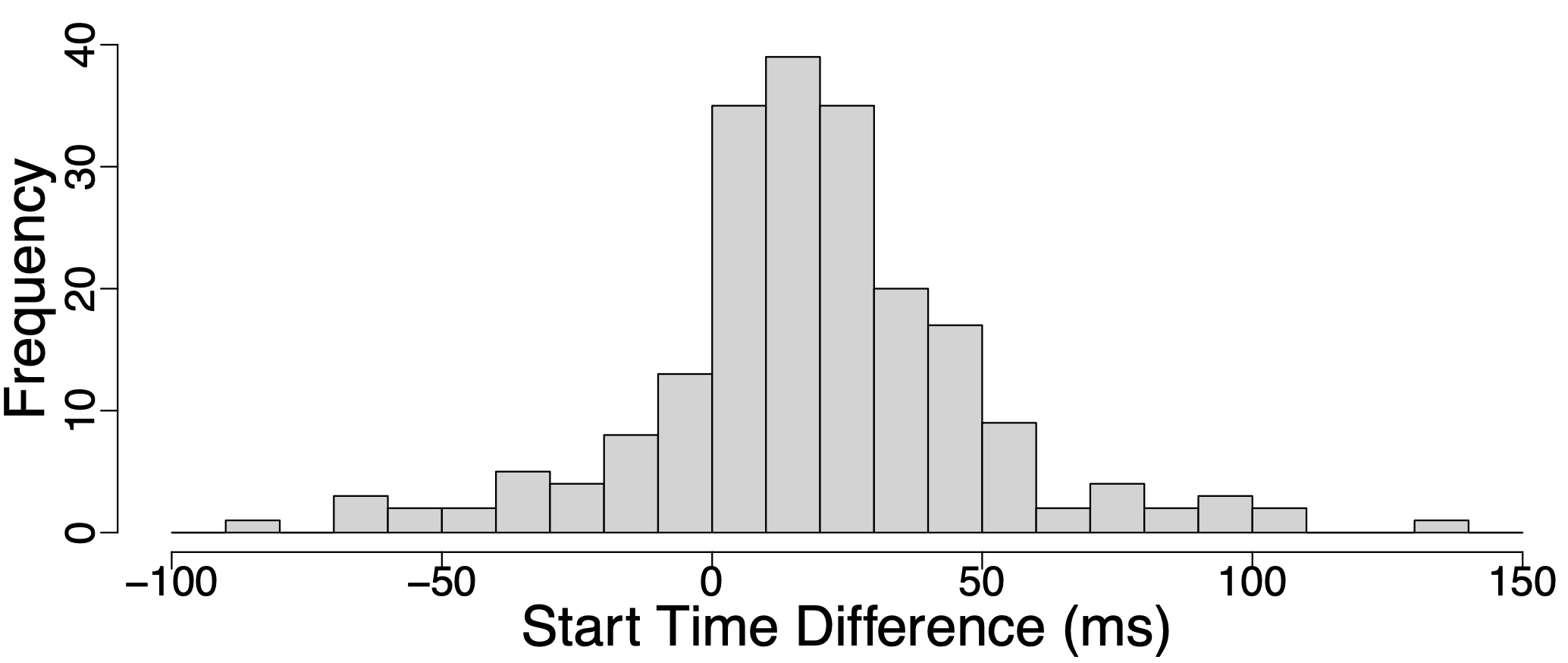}}
\caption{Histogram of detection time difference between SWRs detection and filter probability distribution prediction. Events whose decision state probability passes the threshold $0.99$ with duration of high probability more than $15$ ms are counted.}
\label{fig6}
\end{figure}

Fig. \ref{fig6} shows a histogram of the difference in start times between the noncausal ripple detector and our causal filter estimates. As expected, there is typically a delay for the causal filter detection times, since the noncausal estimator is able to use future information to predict events in advance and is designed to produce conservatively large ripple period estimates. However, this delay tends to be small compared to the average duration or a ripple-replay event. The mode of the detection delay is about $10$ ms, and about $99\%$ of the delays are shorter than $100$ ms. 

\begin{figure}
\centerline{\includegraphics[width=8.8cm, height=5.3cm]{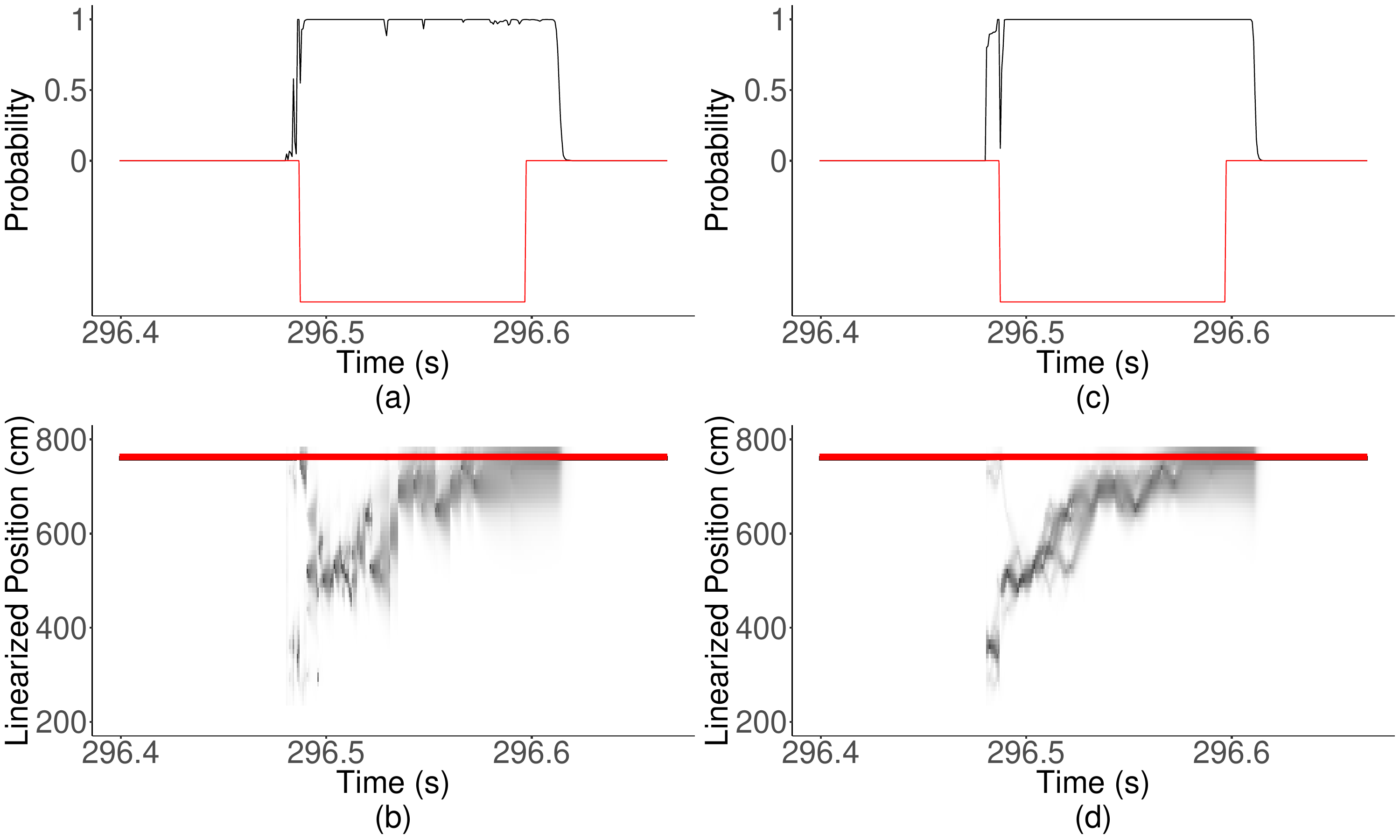}}
\caption{Filter and smoother comparison. (a) Filter probability of a replay event at current time. The red line is result from noncausal SWRs detection. (b) Heat plot of filter probability distribution for  semi-latent continuous variable. The red line is the observed linearized position, and the shadow area is the distribution. (c) Smoother probability of a replay event at current time. The red line is result from noncausal SWRs detection. (d) Heat plot of smoother probability distribution for semi-latent continuous variable. The red line is the observed linearized position, and the shadow area is the distribution.}
\label{fig7}
\end{figure}

A comparison between the filter and the smoother probability distribution for real data is shown in Fig. \ref{fig7}. Fig. \ref{fig7}(a) shows the filter probability of a replay event at each time step. The red line is the estimate from the noncausal ripple detector. Fig. \ref{fig7}(b) shows a heat plot of the filter probability distribution for the semi-latent continuous variable $x_k$. The red line is the observed linearized position, and the grayscale region is the filter distribution. Fig. \ref{fig7}(c)-(d) show the equivalent estimates for the smoother, using future data as well. While the estimates are a bit smoother and certain than the filter estimates, there is minimal difference in the estimated ripple-replay interval. 

\section{Discussion}
In this work, we developed a state-space framework that combines information from populations of spiking neurons with rhythmic neural field activity, which does not require a separate spectral estimation step, and captures switching between multiple brain states that contain different rhythmic and spiking content. The framework uses multiple latent processes to capture the state of a set of biological and behavioral signals that influence neural spiking, the state of a set of internal oscillators that influence the observed neural fields, and a discrete switching state that reflects different patterns of activity in the field and spiking data between exploration and replay periods. Critically, this framework allows for estimation of the discrete brain state with only past and current data, a critical step to real-time estimation that incorporates information from both spike and LFP data. We applied this framework to the problem of identifying and decoding ripple-replay events in rat hippocampus. In this context, the framework differentiates between non-replay states, which contain spiking activity that reflects the actual movement of the rat and LFP data dominated by theta rhythmicity, and ripple-replay states, which contain spikes that reflect nonlocal representations of space and LFPs with enhanced low frequency and ripple frequency activity. 

Earlier methods for identifying ripple-replay events typically involved a multi-step process, where each step required computations that were noncausal or would not be computable in real-time settings. Moving from spectral estimators of oscillations in field data to the state-space oscillator model eliminates the need for spectral estimates that are typically noncausal or else provide lagged estimates. Moving from sorted spiking models to clusterless spiking models eliminates the need for spike sorting, which is typically too time intensive to permit real-time decoding. In fact, Ciliberti et al. demonstrated that real-time replay detection was achievable using clusterless spiking models for closed-loop experiments \cite{b89}. While this methods differs from the one we proposed in not using a latent model structure to identify nonlocal spatial trajectories and to integrate information from LFP rhythms, it does demonstrate that one of the major challenges for closed-loop implementation of our methods - real-time identification of spikes and waveforms - can be achieved with advanced hardware and software methods.

Recent research into associations between LFP ripple and spiking activity has highlighted the importance of integrating information from each of these data streams \cite{b56}, \cite{b90}, \cite{b91}. Our combined model structure not only makes both sources of information available for analyses in causal manner, but also dynamically computes the contribution of each data source to the final state estimate. In our simulation scenarios, compared to other methods, our approach demonstrates superior accuracy and swiftly detects ripple-replay events using only the neural signal information available up to current moment. This is achieved by fully leveraging information from both LFPs and spikes, seamlessly integrating them through the state-space model. In our analysis of real spiking and LFP data from rat hippocampus, we found that this state-space approach was able to detect the overwhelming majority of events detected by prior noncausal methods that used only LFP data, and found many additional periods of nonlocal neural spiking that were not detected by previous methods. Of those replay events detected by both methods, our estimator detected the majority within $30$ ms of the onset estimated by the prior, noncausal method, despite having no access to the future data. Our methods also provide an explicit estimate of the certainty of each detection, which was not available using the prior method.

One important issue is how the computational complexity of our methods scales as the dimensionality of our state processes (the dimensionality of the position state and the number of oscillators) and the number of LFP channels increases. The complexity of the iterative update step in (\ref{eq4}) arises from integrating over the values of the state processes $x_{k-1}$ and $\vec{Z}_{k-1}^{(1:D)}$, and from the multiplication observation distributions for $\vec{Y}_{k}$ and $\Delta N^{(1:E)}_k$. This complexity is similar to other state-space filters, such as the Kalman filter, which for naive implementations scale quadratically with the dimension of the state process and polynomial with exponent smaller than $3$ for the dimension of the observation process \cite{b73}, \cite{b74}. There are multiple approaches to keep the dimensionality of these variables and thereby the complexity of the computations in line for real-time applications.  Linearization is one of the most commonly used methods to help improve computational efficiency in decoding spatial content during rodents' movement \cite{b8}, \cite{b13}, \cite{b14}, \cite{b31}, \cite{b32}. These methods have already been used to achieve real-time decoding of hippocampal replay content \cite{b62}. High-performance linear algebra libraries, like RcppArmadillo \cite{b59} or Intel Math Kernel Library \cite{b60}, can also help preserve the computational efficiency of these methods. We implement the iteration in (\ref{eq4}) using particle filter methods, and the computation for each particle can be computed independently of the others using parallel computing methods. High-performance CPUs (eg. Intel Core i9-13900K (24 cores, 32 threads)) and/or GPUs (eg. NVIDIA RTX 3080/3090) can accelerate parallel computation \cite{b70}, \cite{b71}, \cite{b72}.

There are a number of ways this model can be extended. Here we assumed that the model parameters could be inferred separately from the estimation of the state process. This is not an unreasonable assumption in the case of hippocampal replay, where place field properties and features of the theta rhythm can be estimated during periods of active exploration. In other neural estimation problems, it may not be as simple to separate model encoding and decoding. For such cases it will be necessary to develop methods to estimate the model parameters and state processes simultaneously. Potential approaches to solve this problem include the EM algorithm \cite{b19}, \cite{b42}, variational Bayes methods \cite{b43}, \cite{b44}, or fully Bayesian approaches \cite{b45}, \cite{b46}. Additionally, the form of the models used to decode hippocampal data are of moderate dimension. The clusterless spiking models used a four-dimensional mark based on the peak height of each spike on each of the four channels of the recording tetrodes, and only a small subset of LFPs were used for decoding. More generally, clusterless decoders can be applied to data from high density probes which will lead to much higher dimensional marks and more channels from which to record local fields. It is still unclear the extent to which these models will scale to high-dimensional settings, and the integration of these methods with dimensionality reduction tools may be necessary. 

In addition, for simplicity of computation and interpretation, we assumed that the semi-latent variable $x_k$ follows a random walk during replay state. This assumption has been used successfully in a number of previous analyses of replay content \cite{b8}, \cite{b13}, \cite{b14}. However, this is almost certainly a simplification of the structure of replay trajectories. Random walks lack biological realism, as hippocampal replay often reflects goal-directed behavior and non-random patterns related to prior experiences or future planning \cite{b75}, \cite{b76}. Additionally, real replay paths exhibit temporal order, spatial continuity, and sequence compression, which random walks fail to capture \cite{b77}, \cite{b78}. Third, random walks ignore environmental constraints, such as boundaries and obstacles that shape replay paths \cite{b80}, \cite{b81}. Fourth, random walk models do not match experimental data, which shows biased, non-uniform replay patterns associated with specific trajectories or high-value paths \cite{b84}, \cite{b82}, \cite{b83}. Finally, replay often involves non-spatial experiences, which would not be expressible as a random walk \cite{b93}, \cite{b94}, \cite{b95}. Alternate state update models can be incorporated into our method to reflect better spatial or non-spatial behaviors represented during replay events. For example, a drift diffusion model can be more suitable to describe the rodents' behavior during hippocampal replay \cite{b63}.


Here we focused on the problem of detecting ripple-replay events in rat data, but another important application is SWRs detection in human recordings. There are still multiple challenges for applying these methods to human data. To begin, behavioral states in humans and animals are characterized differently. While the behavioral correlates of SWRs in rodents are well characterized, quantitative description of correlations in humans is lacking due to technical constraints or clinical limitations \cite{b56}, \cite{b69}. Second, human hippocampal ripples may be confounded by pathological high-frequency oscillations linked to epilepsy, making it difficult to distinguish physiological ripples from pathological activity \cite{b85}. In addition, reference electrode placement in humans is limited by clinical constraints. When multiple recording electrodes are used, electromyography artifacts are synchronously recorded on most of them \cite{b56}, \cite{b67}, \cite{b68}. More importantly, even when electrodes are confidently located in the CA1 pyramidal layer, frequency band, duration, and amplitude thresholds for detecting hippocampal SWRs vary widely across rodent, non-human primate, and human data \cite{b56}.

While the focus of this paper is in applying the model framework to the problem of detecting and decoding ripple-replay events, this model structure is widely applicable to neural estimation problems that include population spiking and coherent rhythmic neural activity. For example, this framework could further elucidate the changes in rhythmic spiking activity in the striatum that underlie volitional movement \cite{b8}, \cite{b14}, \cite{b39}. More generally, this model framework could be applied to problems examining communication through coherent activity between brain areas where spikes and LFPs are simultaneously recorded \cite{b8}, \cite{b40}, \cite{b41}.

\section{Conclusion}

At its heart, the state-space modeling approach offers a natural framework for integrating information from multiple observed signals to draw inferences about dynamical neural processes \cite{b1}, \cite{b47}, \cite{b48}, \cite{b49}, \cite{b50}. This work expands the space of models to include causal descriptions of multiple neural oscillations. This makes it possible to identify in real-time neural processes that are characterized by combinations of population spiking and observed rhythmic activity. More broadly, it brings us closer to a universal framework that can combine information from all observed signals in an experiment to understand cognitive processes more completely.

\section*{Acknowledgment}
This work was supported by research grants from the Simons Collaboration on the Global Brain (542971, NC-GB-CULM-00002730) and the NIH (RF1MH130623). We thank Dr. Loren Frank, Dr. Eric Denovellis and Dr. Anna Gillespie for providing data and for many helpful discussions related to these methods.

\section*{Declaration}
The authors declare that they have no conflict of interest. The paper is currently under review for publication with IEEE Transactions on Biomedical Engineering (TBME).

\end{document}